# Magnetic evaluation of a developed permanent magnet undulator as a linac-based THz coherent radiation source

Ali Ramezani Moghaddam[a]

[a]Physics and Accelerators Research School, Nuclear Science and Technology Research Institute, Tehran, 14155-1339, Iran

ABSTRACT: Due to the various applications of THz sources, we developed a pure permanent magnet (PPM) undulator with 8 periods and min/max values of gap equal to 15/25, to use at the end of our current 10 MeV linear accelerator (linac) in order to produce THz coherent radiation. We used RADIA for simulation and calculation of our magnet and B2E code for calculation of its radiation harmonics. After fabrication, we also measured the magnetic field via high precision Hall probes. Calculations and measurements are in good agreements. These calculations show that it is possible to reach photon energies from 3 meV to 23 meV in the THz region.

## 1. INTRODUCTION

THz-Band of electromagnetic spectrum, 0.1-10 THz, has interesting properties for many applications [1-4]. Many different disciplines such as ultrafast spectroscopy, semiconductor device fabrication, condensed matter physics, and bio-medical imaging involve the recent development of THz technology [5-10]. THz beams transmitted through materials can be used for study material characterization, layer inspection, and as an alternative to X-rays for producing high resolution images of the interior of solid objects [11-13].

Nowadays, the generation and detection technology of THz radiation is improving rapidly. There are variable sources of THz radiation with specific properties [14-17]. Among them, accelerator-based THz light sources are well-known as a source with high flux, power and reliable performance [18-21]. There are two approaches to reach THz radiation by using accelerators: coherent transition radiation and the undulator based coherent radiation. Both of them need the electron bunch length less than 1 picosecond (ps) to produce high coherent THz radiation [22-25].

We have a domestic RF linear accelerator with electron energy of 8-15 MeV [26-27]. We are going to use it for generating coherent THz radiation. It is capable of producing pico and under-pico second bunch length [28]. Table 1 shows the measured parameters of the electron beam of this linac.

**Table 1:** Measured properties of electron beam at the linac [26].

| Property | Value |
| --- | --- |
| Resonance frequency | 2998 MHz |
| Repetition rate | 10 Hz |
| RF pulse length | 6 µs |
| Maximum RF power | 6 MW |
| Bunch charge | 0.1 nC |
| Bunch length | 200 fs – 5 ps |

Nowadays, with the development of permanent magnet fabrication technology, pure permanent magnets (PPM) have been used widely in various industries [29]. Especially, an enormous number of undulators around the world are PPM undulators as a synchrotron or FEL radiation source [30]. PPM undulators have some advantages against the electromagnetic undulators. For instance, permanent magnets are hard ferromagnetic material and have linear behavior whereas in electromagnetic undulator the poles material are soft ferromagnetic and have strong non-linear behavior [29,30]. This property is vital in magnetic field tuning by changing the gap value since in PPM undulator, the final magnetic field is the sum of the fields from each magnet [31]. Also, the shimming process of the undulator and magnetic field correction in PPM undulator is simpler, especially, when the undulator is used at higher harmonics [32,33]. In PPM undulators, there is no need of a power supply which helps us to save energy and to eliminate cooling system. It is possible to reach the small period and high magnetic field values with PPMs. It is not a hard work to adjust the undulator gap value by using a simple and reliable mechanical design. For these mentioned reasons, we chose the PPM structure for our undulator. In this paper, we explain the steps of the design and calculations of the PPM undulator for producing THz radiation from our linac.

## 2. SPECIFICATION OF UNDULATOR MAGNET

Figure 1 shows the main geometrical parameters of a PPM undulator. The on-axis n[th] harmonic photon wavelength $\lambda_n$ is related to undulator period, $\lambda_u$, and the undulator parameter, $K = 0.934\lambda_u B$ as in the following equation [34], ($\gamma$ is the relativistic Lorentz factor).

$$\lambda_n = \frac{\lambda_u}{2n\gamma^2}\left(1 + \frac{K^2}{2}\right) \tag{1}$$

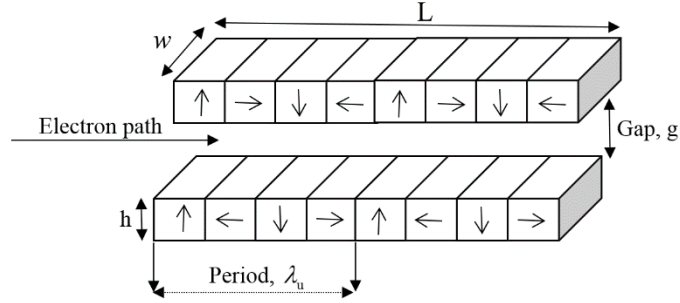

**Figure 1.** A Schematic view of a PPM undulator fabricated by two permanent magnet arrays.

In order to tune the photon energy, the more practical way is the magnetic field adjustment and using BPMs [35-36]. In electromagnetic and hybrid undulators, the magnetic field behavior depends strongly on the pole material which is fabricated from soft magnetic material, e.g. vanadium permendur [37]. For this reason, the shimming and optimization of final field is a serious challenge for this kind of undulators. Generally speaking, the PPM undulators are very robust and reliable devices to tune the magnetic field by varying the undulator gap value [38]. Fortunately, there is an analytical relation among the on-axis peak magnetic field, $B_0$ and the gap, $g$ and magnet block height, $h$ as [38].

$$B_0 = 1.682 B_r \lambda_u \left(1 - e^{-2\pi h/\lambda_u}\right) e^{-\pi g/\lambda_u} \tag{2}$$

Where, $B_r$ is a remanence field in the magnet. We can reach to 96% of maximum field by choosing $h = \lambda_u/2$. The grade of magnet material determines two important factors of the magnet, i.e. the remanence, $B_r$, and the coercivity, $H_{ci}$. The remanence field specify the peak field and the coercivity specify the resistance of the magnet against the demagnetizing field. The source of demagnetizing fields is the nearby magnets at upper and lower jaws and must be calculated by RADIA [39].

According to equations (1) and (2), it is possible to reach almost 3 meV (0.7 THz, 427.5$\mu m$) and 9 meV (2.12 THz, 142.5$\mu m$) for the 1[th] and 3[rd] harmonics, respectively, with the gap of 15 mm, and 7.7 meV (1.86 THz, 160$\mu m$) and 23 meV (5.6 THz, 53 $\mu m$) for the 1[th] and 3[rd] harmonics, respectively, for the gap of 25mm by choosing 48 mm as the period length. We chose NdFeB material with the grade of SH35 for our purpose. The remanence ($B_r$) and intrinsic coercivity ($H_{ci}$) of this grade are 1.18 T and 1592 kA/m, correspondingly. The optimum value for undulator deflection parameter (K-value) to generate the 1[th] and 3[rd] harmonics is 1. As the Fig. 2 shows, at K=1, the harmonics intensities ($\propto F_n$) are at maximum value and by increasing the K-value, the higher harmonic numbers are produced and these profitless harmonics will generate high power loss on the target. For some economic reasons, we had to use our available permanent magnets which gives us final K-value of 3.36 for the maximum peak field of 0.75 T at the minimum gap of 15 mm. A robust mechanical structure was designed for field adjustment. However, it helps us to change the harmonics energy as well as the K-value by changing the gap value.

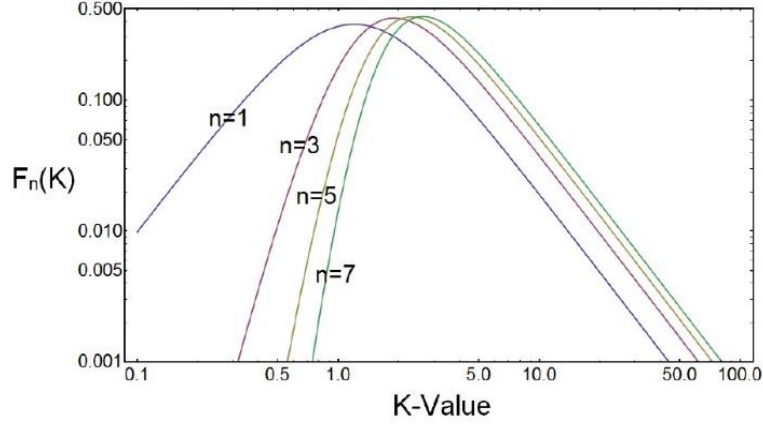

**Figure 2.** Different harmonic intensities behavior versus the K-value.

The total length of the undulator should not be larger than 2 m, which depends on the desired flux and coherency as well as the available free space at the laboratory hall. Due to these limitations, we chose 41 periods in our undulator. The minimum value for the undulator gap depends on the vacuum chamber dimensions [40-41]. Both of attractive magnetic forces between upper and lower jaws as well as the good field region (GFR) of the undulator depend on the magnet block width (*w* in Fig. 1). This value must be optimized to reach the desired GFR and minimum attractive force and also require lower magnetic material to save money. We can derive the rms-error of the peak field of the undulator, $\sigma_B$ versus the rms-error of the gap, $\sigma_g$ and rms-error of the remanence, $\sigma_{B_r}$ as the following equation [38],

$$B \propto B_r e^{-\pi g / \lambda_u} \Rightarrow \frac{\sigma_B}{B} = \sqrt{\left(\frac{\sigma_{B_r}}{B_r}\right)^2 + \left(\frac{-\pi}{\lambda_u}\sigma_g\right)^2} \qquad (3)$$

Where *g* is the gap, $\lambda_u$ is the period of the undulator and B is the on-axis peak magnetic field. $\sigma_g$ is twice the deformation of a single girder of undulator. To minimize the fluctuations of the peak field, one must keep $\sigma_g$ as small as possible. In order to determine our desired magnetic field errors to reach appropriate intensity for the 1th and 3rd harmonics, the Kincaid model has been used. In this model, an ideal sinusoidal magnetic field with random errors on individual poles is described by a series of half-sinusoids [42]. First, we should generate some semi-real magnetic fields with a specific rms-field error and calculate their effects on the radiation harmonics intensity. We can use rms-field error to calculate rms-phase error to qualify the undulator magnetic field. The field errors in the real magnetic field of the undulator are due to the inhomogeneity of magnetic material in the permanent magnets and mechanical errors. This first source of field error is investigated by using the Helmholtz coil to measure the quality of the permanent magnets. Figure 3 shows the rms field error effects on the harmonic intensity for some initial harmonics. We can conclude from Fig. 2 that we need almost 1.5% rms field error to reach higher than 80% of ideal intensity for the 1th and 3rd harmonics. Our measurements by Helmholtz coil shows that the magnet block magnetization has a 1% error in magnitude and 1.5% in direction. According to equation (3) the relative rms field error is 1.5%, then for the maximum rms error for the gap value we have:

$$1.5 \times 10^{-2} = \sqrt{(10^{-2})^2 + \left(-\pi \frac{\sigma_g}{48(mm)}\right)^2} \Rightarrow \sigma_g \approx 171 \mu m \qquad (4)$$

Based on our calculation which is shown in equation (4), maximum rms error for the gap value must be $171 \mu m$.

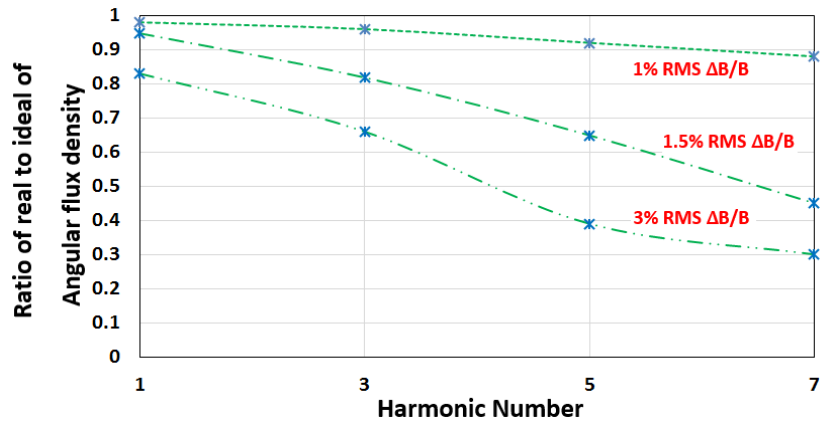

**Figure 3.** Real to ideal ratio of harmonics intensity for three RMS field errors.

## 3. MAGNETIC AND MECHANICAL DESIGN OF UNDULATOR

RADIA was used for 3D simulation of the undulator magnet structure [43]. The symmetrical field design was used. In the case of symmetrical design, the on-axis magnetic field at the center of the undulator is at its maximum value [44, 45]. In order to get rid of the first and second field integral at these types of undulators, one can add half width magnets at the both ends of each section (See Fig. 4) [46]. Figure 5 shows the on-axis magnetic field at the minimum and maximum values of the gap, i.e., high and low K-values, respectively. Figure 6 depicts the effects of end section of the undulator on the beam trajectory at the central surface of the undulator. The end section is important since it minimizes the first and second field integrals that specify the angle and position of beam, correspondingly. Elimination of end sections can induce some phase errors. The non-zero values for field integrals and phase errors dramatically reduce the intensity of harmonics and also can change their energy. In circular synchrotron ring, control of these errors are very critical for beam stability. In THz or IR undulator sources, the beam will be dumped after passing through the undulator.

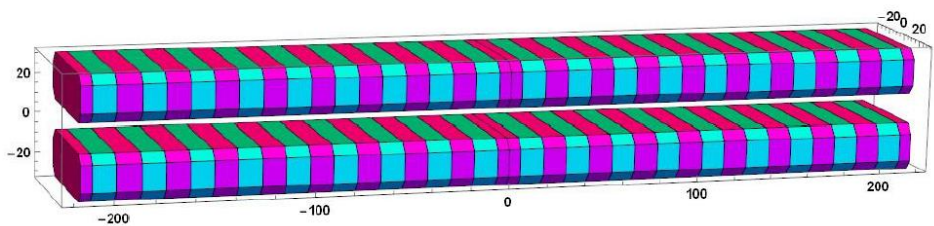

**Figure 4.** 3D model of prototype undulator with 8 periods and gap value 15mm.

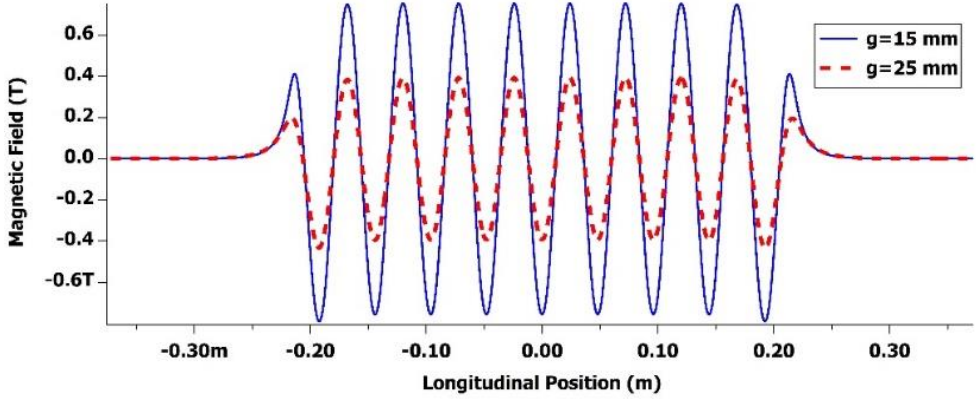

**Figure 5.** Undulator on-axis magnetic field for Min and Max gap value.

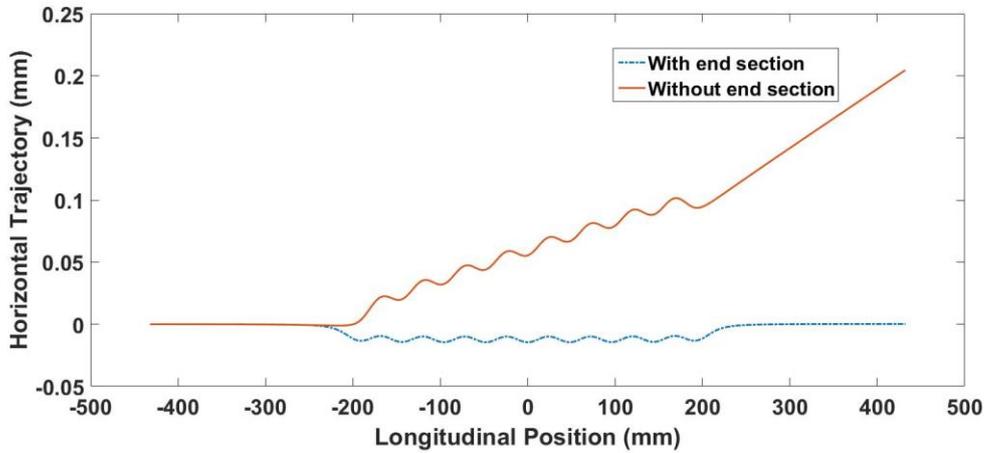

**Figure 6.** Horizontal electron trajectory with and without undulator end-sections.

There are two main challenges in the mechanical design. The first is to control the attractive magnetic force between the upper and lower jaws for different values of the gap and the second challenge is assembling of the permanent magnets close together until they touch each other. Our goal is to reach our necessary tolerances economically, as far as possible. Aluminum has been chosen as the jaws material. In many long and strong undulators, the stainless steel has been used for this purpose which is more expensive than aluminum and also after machinery, the steel magnetic properties has been changed and it is necessary to do some annealing process for demagnetization. For assembling of the permanent magnets close together, a simple mechanism has been designed and fabricated by using a long arm screw and an aluminum guider. The strong magnetic force acts on the permanent magnets where assembled in two jaws. Figure 7 shows the vertical and longitudinal magnetic force on the magnets in the upper jaw from all magnets mounted on the undulator at the minimum gap value (=15mm).

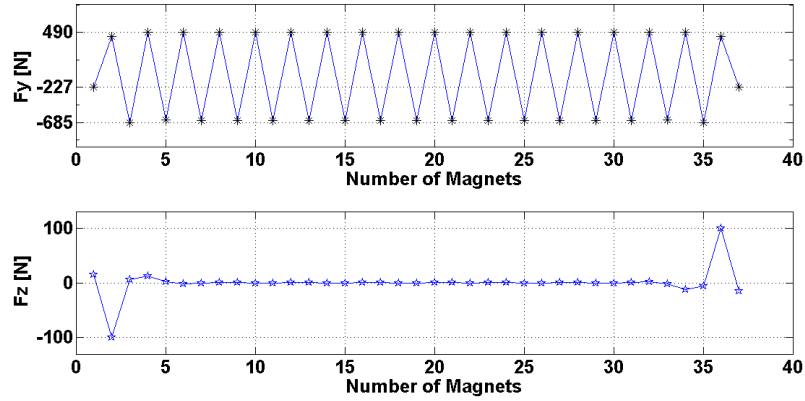

**Figure 7.** The vertical, Fy and longitudinal, Fz magnetic force acting on the upper girder's magnet in the minimum gap=15mm calculated by RADIA [43].

With the vertical magnetization is in the outside direction of the jaw and in the case of the magnets with the horizontal magnetization direction, the magnetic force pushes the magnets inside the jaw. For the lower jaw, the magnetic forces act in the same way of the upper one. The magnet blocks have been tightened by using the non-magnetic steel screws.

The attractive magnetic force between upper and lower jaws is an important factor to design the gap variation mechanism. This force is calculated by RADIA and is equal to 3600 N [43]. The gap value control system is shown in Figure 8.

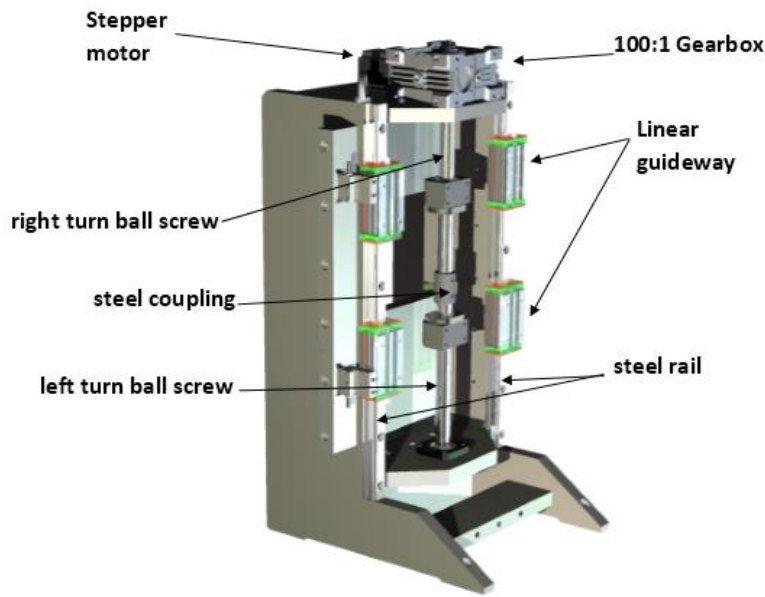

**Figure 8.** Undulator driving mechanism to control the gap value.

## 4. ESTIMATION OF UNDULATOR RADIATION

We calculated the radiation harmonics using B2E codes [47]. Via these codes, we can use the real magnetic fields to calculate the radiation harmonics using far field approximation. The ideal magnetic field from RADIA simulation and real magnetic field from Hall probe sensors were imported to the code and the radiation angular flux density was calculated and compared. At the end, radiation angular flux density of the final undulator with 41 periods was calculated. Figures 9 and 10 show the THz radiation harmonics from the undulator with the gap of 15 mm and 25 mm, respectively, and compare the real and ideal harmonics together. Figure 11 shows the harmonics intensity from a 41 period undulator. The harmonics FWHM and their intensity could be compared with the 8 period prototype undulator.

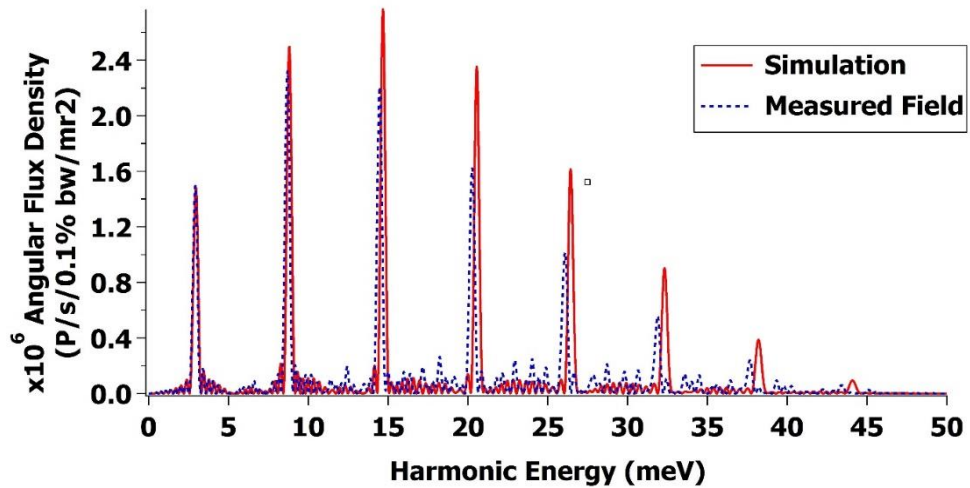

**Figure 9.** Real and simulated THz radiation harmonics from 8 period's undulator at gap 15 mm.

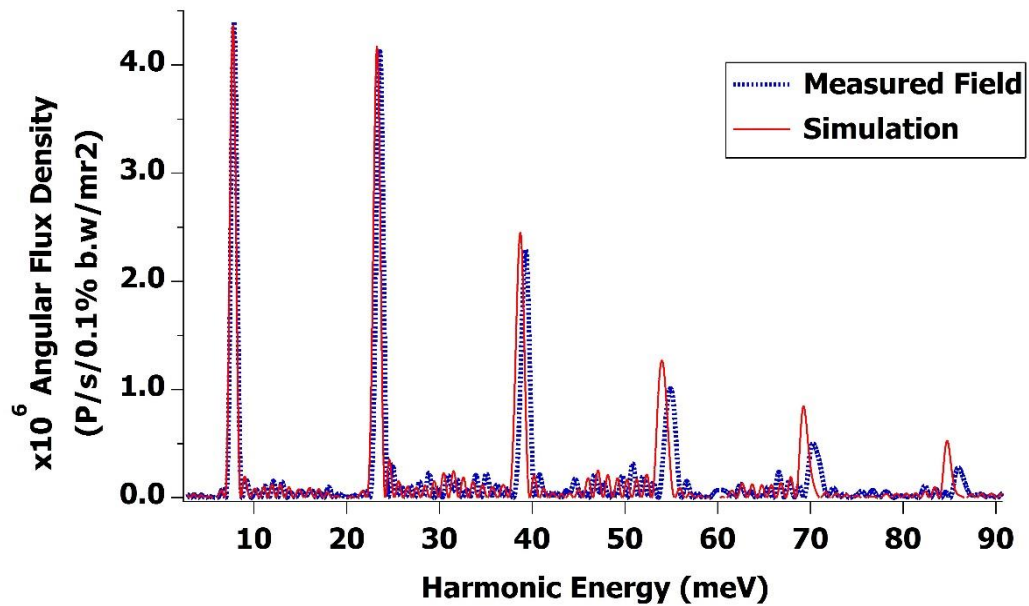

**Figure 10.** Real and simulated THz radiation harmonics from 8 period's undulator at gap 25 mm.

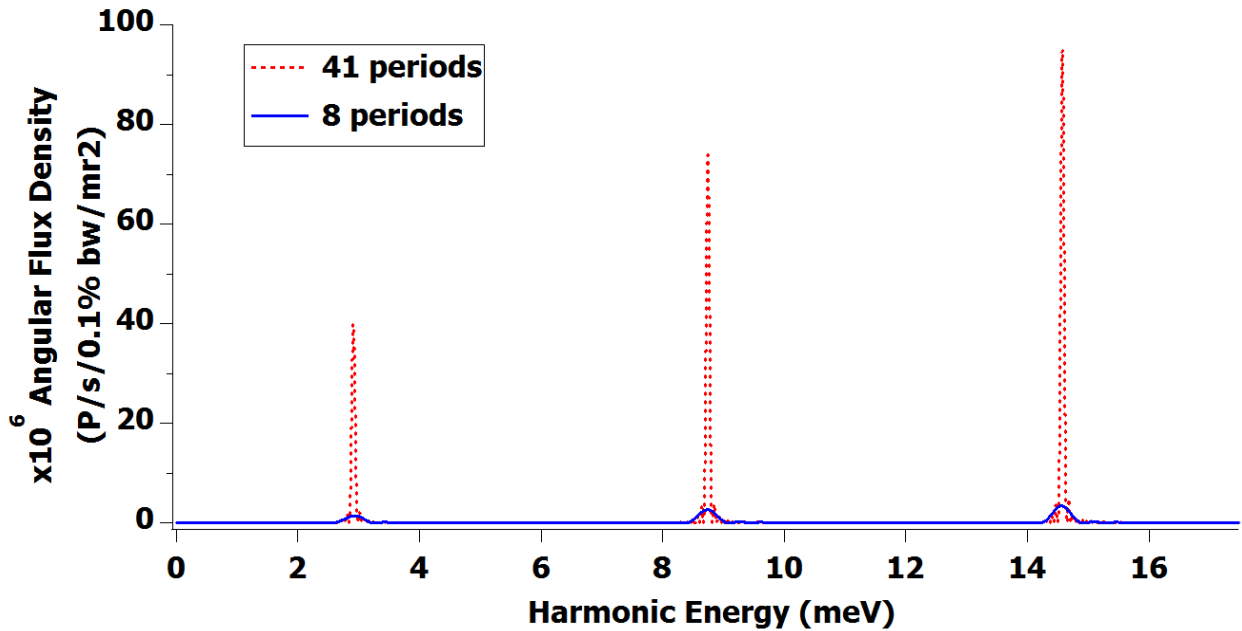

**Figure 11.** THz radiation harmonics from 8 periods and 41 periods' undulator at gap 15 mm.

## 5. Conclusion

A prototype pure permanent magnet with 8 periods and min/max value of gap equal to 15/25 has been developed for producing THz radiation by a 10 MeV linear accelerator. The K value of the undulator at minimum gap of 15 mm is 3.36. For economical purposes, we used available permanent magnets in our lab. We studied tolerances in order to have an accepted range of field deviation. Based on our calculations, maximum rms error for the gap value must be $171 \mu m$. We used RADIA to model 3D structure of the undulator and B2E to calculate the real and ideal harmonic radiation [43,47]. A Helmholtz coil was used to measure the magnetization at each magnet block to insure the desired tolerances. At the end, calculation results showed a good agreement with measurements. The undulator can generate 3 meV to 23 meV photon energies in THz region.